\documentclass[journal,draft,onecolumn,12pt]{IEEEtran}
\usepackage{amsfonts,color,morefloats}
\usepackage{amssymb,amsmath,latexsym,amsthm}
\usepackage{float}
\usepackage{algorithm}
\usepackage{algorithmic}
\usepackage{multirow,tabu}
\usepackage{nicematrix}
\usepackage{amsmath}
\usepackage{multirow, bigdelim}
\usepackage{enumitem}
\usepackage{hyperref}
\newcommand{\Bn}{\mathcal B_n}
\newcommand{\RM}{\operatorname{RM}}
\newcommand{\nlr}[1]{\operatorname{nl}_{#1}}

\newcommand{\wt}{\operatorname{wt}}
\newcommand{\dist}{\operatorname{dist}}

\newcommand{\rank}{\operatorname{rank}}
\newcommand{\dd}{\delta}

\newcommand{\F}{{\mathrm{GF}(2)}}

\newtheorem{theorem}{Theorem}
\newtheorem{lemma}[theorem]{Lemma}
\newtheorem{proposition}[theorem]{Proposition}
\newtheorem{corollary}[theorem]{Corollary}

\definecolor{myred}{RGB}{255,0,0}

\begin{document}

\title{The Covering Radius of the Reed--Muller Code \texorpdfstring{$\RM(3,8)$}{RM(3,8)} is 56\thanks{
  This work was supported by the National Natural Science Foundation of China (Grant Nos. 62372247 and 12441103), and by open research fund of State Key Laboratory of Cyberspace Security Defense (No. 2025-MS-03).
}}

\author{Zhihang Zhao,\thanks{Zhihang Zhao is with the School of Computer Science, Nanjing University of Posts and Telecommunications, Nanjing, 210023, China. Email: 2310273018@qq.com.} 
\and Yansheng Wu*\thanks{Yansheng Wu is with the School of Computer Science, Nanjing University of Posts and Telecommunications, Nanjing, 210023, China and the State Key Laboratory of Cyberspace Security Defense (Institute of Information Engineering, Chinese Academy of Sciences, Beijing 100085), China. Email: yanshengwu@njupt.edu.cn.}
}

\date{\today}
\maketitle

\begin{abstract} 
We prove that the covering radius of the third-order Reed--Muller code in eight variables is fifty-six. The proof separates even- and odd-weight inputs by moment-controlled slicing. On the even side, a transport invariant on quadratic-label stars and consistency of singleton restrictions give a compatible cubic correction. On the odd side, symmetry of the third moment yields a covariance-rank obstruction. Its rank-two exception is a quartic plus three points of a two-plane; the zero-second-moment case is a prescribed-point perturbation of a quintic. We prove that every such point perturbation has third-order nonlinearity at most fifty-five by a marked derivative identity and complete restriction-code covers. Consequently the maximum among exact-degree-eight functions is fifty-five. The structural reductions are proved analytically, and complete finite certificates verify the explicitly stated local covering and compatibility assertions. The matching lower bound follows from the published quartic relative radius.
\end{abstract}


\section{Introduction}
\label{sec:introduction}


{
For \(0\le r\le n\), the \(r\)-th order Reed--Muller code, denoted by \(\mathrm{RM}(r,n)\), consists of all Boolean functions in \(n\) variables whose algebraic degree is at most \(r\). The covering radius of a code is defined as the smallest integer \(\rho\) such that every vector in the ambient vector space is within Hamming distance \(\rho\) of some codeword.  Determining the covering radius \(\rho(r,n)\) of the \(r\)-th order Reed--Muller code \(\mathrm{RM}(r,n)\) is one of the most extensively studied yet persistently elusive metric parameters in coding theory, which is also  a challenging problem that has received widespread attention from many scholars.
Progress has been uneven: exact values exist only in a narrow parameter regime, while for general \(r,n\) the known upper and lower bounds often remain separated by a substantial gap. Here we just list some of them as follows:

\begin{itemize}
    \item The earliest exact result for \(r\ge 2\) is due to Schatz, who proved \(\rho(2,6)=18\). Schatz's survey of covering radius results also systematically collected the known bounds and exact values for Reed--Muller codes available at that time \cite{schatz1981}. 
    
    \item The case \(r=2\) then remained essentially frozen at \(n=6\) for nearly four decades. For \(\mathrm{RM}(2,7)\), the covering radius was long known only to lie in \([40,44]\); intermediate work narrowed the upper bound to \(42\), but the exact value remained open until Wang proved \(\rho(2,7)=40\) in 2019 \cite{Wang2019}. Wang's argument identifies \(\mathrm{RM}(2,7)\) as having the same covering radius in \(\mathrm{RM}(3,7)\), and the proof proceeds through a careful analysis of affine equivalence classes of \(6\)-variable Boolean functions.

    \item  The third-order case has seen more recent activity. For many years the best known bounds for \(\mathrm{RM}(3,7)\) placed its covering radius between \(20\) and \(23\). Gao, Kan, Li, and Wang proved in 2023 that \(\rho(3,7)=20\), closing this gap \cite{GaoKanLiWang2023}. Gao~\cite{Gao2026} determined the quartic relative radius in eight variables, giving the lower bound fifty-six. These results give the interval
$$
56\le\rho(3,8)\le60.
$$ At the fourth order, Gillot and Langevin established in 2024 that \(\rho(4,8)=26\), using an effective version of the derivation-lift invariant to classify the quotient \(\mathrm{RM}(6,8)/\mathrm{RM}(4,8)\) \cite{gillot2024}. Dougherty, Mauldin, and Tiefenbruck had earlier determined the covering radius of \(\mathrm{RM}(m-4,m)\) in \(\mathrm{RM}(m-3,m)\), recovering \(\rho(4,8)=26\) as a special case \cite{dougherty2022}.

    \item For \(r=1\), the covering radius coincides with the maximum nonlinearity of Boolean functions. Patterson and Wiedemann constructed highly nonlinear functions for \(n=15\) and related parameters \cite{patterson1983}, but exact values for odd \(n>7\) remained inaccessible for a long time. The case \(n=9\) was settled only through a computer-assisted proof establishing \(\rho(1,9)=242\), following earlier constructions of \(9\)-variable functions with nonlinearity \(242\) that gave the matching lower bound \cite{kavut2010}. Exact values of \(\rho(1,n)\) are still unknown for general odd \(n\).
\end{itemize}

}



Inspired by the above results, we determine the value in eight variables in this paper.

\begin{theorem}
\label{thm:global}
The covering radius of the Reed--Muller code $\RM(3,8)$ is
$
\rho(3,8)=56.
$
\end{theorem}

{
Our proof uses a degree-sensitive slicing argument. Its first ingredient is a prescribed-point theorem: for every background of degree at most five, flipping any prescribed point gives third-order nonlinearity at most fifty-five. A quartic derivative reduces this assertion to a restriction of the seven-variable cubic code. The marked point changes the coordinate set by one deletion or one addition, and an exact weight identity retains the same correction on both slices. This gives a nearest-residual support property which an ordinary relative upper bound does not supply.

The second ingredient treats the moment geometry of higher-degree inputs. For even-weight functions, paired normal planes give four odd six-dimensional slices. A transport invariant on common-point quadratic-label stars supplies compatible corrections of total cost thirty-eight. In the remaining configuration, singleton positions agree on five-dimensional overlaps; graph-restriction rigidity then forces a global cubic remainder. This proves the even upper bound without assuming any sixth-order relative conclusion.

For odd-weight functions, we center the first moment. A hypothetical distance fifty-seven forces coupled seven-variable endpoints nineteen and eighteen on every nonradical normal. The centered covariance of the odd restriction is constrained by symmetry of the third moment. A Gram-rank argument rules out simultaneous vanishing at second-moment ranks four, six and eight. At rank two, the exceptional configuration is a quartic plus three points of a two-plane, which has a direct upper-fifty-five proof. Zero centered second moment leaves a singleton perturbation of a quintic, precisely the prescribed-point problem solved in the first step.

Complete finite verification establishes the local statements required by these reductions: restriction-code covers, local sextic corrections, quadratic-label stars, and the nonzero-covariance endpoint. The classifications of Gillot and Langevin~\cite{GillotLangevin2023} provide their complete domains. Every first-slice coefficient and every relative affine position is retained by the reductions in the text. The computations verify the specified local interfaces, rather than a classification of all eight-variable cubic cosets.

The proof of the seven-variable result in~\cite{GaoKanLiWang2023} organizes concatenated functions by the eleven affine classes of six-variable cubic cosets and then reduces the remaining types. Here the decisive structure is the interaction of moments, compatible slice corrections and a prescribed point. In particular, the covariance-rank argument and the marked derivative identity keep the relative data needed for the final odd-weight exclusion.

The paper is organized as follows. 
Section~\ref{sec:relative} proves the prescribed-point theorem. Section~\ref{sec:sextic} gives the strict local sextic bound, and Section~\ref{sec:r6} develops quadratic-label compatibility. Section~\ref{sec:unified} proves the even-weight bound. Section~\ref{sec:odd-moments} gives the covariance reduction, and Section~\ref{sec:unrestricted} proves Theorem~\ref{thm:global}. The finite obligations and reproducibility are specified in Section~\ref{sec:verification}.

}

\section{Preliminaries and external inputs}
\label{sec:preliminaries}

In this section, we will recall some basic concepts and results on Boolean functions and covering radii. We mainly refer to reference \cite{Carlet2021} for this section of the knowledge.

Let $\Bn$ be the space of Boolean functions in $n$ variables. Addition of Boolean functions is pointwise addition over the binary field. Hamming weights are ordinary integer weights. Every Boolean function has a unique algebraic normal form
$$
f(x_1,\ldots,x_n)
=
\sum_{S\subseteq\{1,\ldots,n\}}
\alpha_S\prod_{j\in S}x_j,
\qquad
\alpha_S\in\F.
$$
The algebraic degree of $f$, denoted by $\deg(f)$, is the number of variables in the highest order term with nonzero coefficient.
For integers $n$ and $r$ satisfying the usual order constraints, define the Reed--Muller code by
$
\RM(r,n)
=
\{f\in\Bn:\deg(f)\leq r\}.
$
The Hamming weight of a vector $x \in \F^n$, denoted by $\wt(x)$, is the number of nonzero coordinates. The weight of a Boolean function $f$, denoted by $\wt(f)$, is the cardinality of the set $\{ x \in \F^n : f(x) = 1 \}$.
The order-$r$ nonlinearity of a Boolean function in $n$ variables is
$$
\nlr{r}(f)
=
\min_{c\in\RM(r,n)}\wt(f+c),
$$ here $f$ is called  the target function.
The unrestricted covering radius of the Reed--Muller code $\RM(r,n)$ is
$$
\rho(r,n)
=
\max_{f\in\Bn}\nlr{r}(f).
$$
The relative radii used below are
$$
R_j=\max_{F\in\RM(j,8)}\nlr{3}(F),\qquad j\in\{4,5,6,7,8\}.
$$
For a point of the input cube, its indicator is
$$
\dd_a(x)=\begin{cases}1,&x=a,\\0,&x\ne a,\end{cases}
\qquad a,x\in\F^n.
$$ A Boolean function is homogeneous if all the monomials of its algebraic normal form are of the same degree. 
We also use the space of homogeneous polynomials
$$
\mathcal H_r(n)=\left\{\sum_{|S|=r}\alpha_S\prod_{j\in S}x_j:\alpha_S\in\F\right\}.
$$

We need the following published results.

\begin{lemma}\cite{GaoKanLiWang2023,Wang2019}
    \label{thm:external-radii}
The following values are available from the literature:
$$
\rho(3,7)=20, \mbox{ and }
\rho(2,7)=40.
$$
\end{lemma}

\begin{lemma}\cite{GaoKanLiWang2023}
\label{thm:external-quartic}
The relative covering radius of the third-order code inside the fourth-order code in eight variables is fifty-six:
$$
\max_{F\in\RM(4,8)}\nlr{3}(F)=56.
$$
\end{lemma}


The complete affine classifications used below are due to Gillot and Langevin~\cite{GillotLangevin2023}. The classification results are used as external completeness and inequivalence inputs; orbit-size sums and checksum comparisons are treated only as consistency checks.

We will also use the elementary parity fact that all codewords in the two seven-variable codes appearing in Lemma~\ref{thm:external-radii} have even weight. Consequently, their distances from odd-weight targets lie in the odd integers. This gives the following immediate refinement.

\begin{lemma}\cite{GillotLangevin2023}
\label{lem:parity}
For every seven-variable Boolean function, the following bounds hold:
$$
\nlr{3}(g)\leq 20,
\qquad
\nlr{2}(g)\leq 40.
$$
If the target function has odd weight, then
$$
\nlr{3}(g)\leq 19,
\qquad
\nlr{2}(g)\leq 39.
$$
\end{lemma}


\section{Cubic corrections through a prescribed point}
\label{sec:relative}

The odd-weight argument will require a cubic residual through a point chosen in advance. We first prove a bound which is uniform in both the first slice and the prescribed point.

\subsection{Quartic derivatives and affine transport}
\label{subsec:relative-derivative}

\begin{lemma}
\label{lem:quartic-derivative}
Every eight-variable function of exact degree five has, after a permutation of its variables, the form
$$
F(x,t)=f(x)+t g(x),\qquad x\in\F^7,\quad t\in\F,\quad\deg(g)=4.
$$
\end{lemma}

\begin{proof}
Choose a variable in a degree-five monomial. Its coordinate derivative has a nonzero coefficient on the complementary four-variable support. No other degree-five monomial contributes to that coefficient, and terms of smaller degree cannot cancel it.
\end{proof}


The classification of Gillot and Langevin~\cite{GillotLangevin2023} gives 68,443 affine classes in the quotient
$$
\RM(4,7)/\RM(2,7).
$$
Exactly 68,431 classes have degree four. Thus, for an affine permutation and one of the representatives, we have
$$
g\circ A=g_i+k,\qquad k\in\RM(2,7).
$$
Apply this permutation to both slices, where a slice means a restriction of a Boolean function obtained by fixing one coordinate (or one linear form). The additional term involving the quadratic is absorbed by a legal cubic correction:
$
tk(x)\in\RM(3,8).
$
The same permutation transports the prescribed point. The first slice is still arbitrary, so no relative position or lower-degree coefficient is lost.

\subsection{The marked restriction-code identity}
\label{subsec:punctured-identity}

For a subset of the seven-dimensional cube, write
$$
\mathcal C_S=\{a|_S:a\in\RM(3,7)\}.
$$
For an arbitrary seven-variable slice difference, choose a quadratic compensation and define
$$
q=g+b,\qquad b\in\RM(2,7),\qquad
Z=\{x:q(x)=0\},\qquad w=\wt(q).
$$
The following identities use no degree condition on the slice difference. They also apply to the cubic-derivative branch of Section~\ref{sec:unified}.

\begin{lemma}
\label{lem:punctured}
For every first slice, the ordinary bound is
$$
\rho(\mathcal C_Z)\le r
\quad\Longrightarrow\quad
\nlr{3}(f+t g)\le w+2r.
$$
For a prescribed point and either position above it, the following bounds hold:
$$
\begin{aligned}
p\in Z,\quad\rho(\mathcal C_{Z\setminus\{p\}})\le R-1
&\quad\Longrightarrow\quad
\nlr{3}(f+t g+\dd_{(p,s)})\le w+2R-1,\\
p\notin Z,\quad\rho(\mathcal C_{Z\cup\{p\}})\le R
&\quad\Longrightarrow\quad
\nlr{3}(f+t g+\dd_{(p,s)})\le w+2R-1,
\qquad s\in\F.
\end{aligned}
$$
\end{lemma}

\begin{proof}
For a cubic function on the seven variables, use the global correction
$$
C(x,t)=a(x)+t b(x).
$$
The corrected slices differ exactly on the support of the compensated derivative. Hence
$$
\wt(F+C)=w+2\wt_Z(f+a).
$$
This proves the ordinary assertion. At a marked point in the zero set, flipping either slice makes the pair contribution one. Approximation on the remaining coordinates gives
$$
\wt(F+C+\dd_{(p,s)})
=w+1+2\wt_{Z\setminus\{p\}}(f+a)
\le w+2R-1.
$$
If the marked point is outside the zero set, its pair contributes zero or two after the flip. On the enlarged coordinate set define the target to equal the first slice on the old zero set and to have value
$
f(p)+1+s
$
at the added coordinate. Its distance from the restriction of the cubic is exactly half the contribution from these coordinates. The other nonzero-derivative pairs contribute one each. Thus
$$
\wt(F+C+\dd_{(p,s)})
=w-1+2\dist(u,a|_{Z\cup\{p\}})
\le w+2R-1.
$$
All corrections used above have degree at most three.
\end{proof}

\begin{lemma}
\label{lem:two-erasure}
If a subset of the seven-dimensional cube omits at least two points, then
$
\rho(\mathcal C_S)\le18.
$
\end{lemma}

\begin{proof}
Extend a target arbitrarily except at two distinct omitted points. Among the four completions, two have even weight. Their first moments satisfy
$$
\sum_x xE_0(x)+\sum_x xE_1(x)=u+v\ne0.
$$
Here the two omitted points are distinct. The first-moment difference is nonzero. At least one even completion therefore has nonzero first moment, and hence exact degree six. The independently proved local bound in Theorem~\ref{thm:sextic} supplies a cubic at distance at most eighteen. Restriction cannot increase distance.
\end{proof}

\subsection{Complete derivative covers}
\label{subsec:relative-certificates}

\begin{proposition}
\label{prop:marked-certificate}
Every exact-quartic derivative class has one of the following certificates.
\begin{enumerate}
\item A quadratic compensation and a restriction-code bound satisfying
$$
w+2r\le54,\qquad\rho(\mathcal C_Z)\le r.
$$
\item For every point, a quadratic compensation and a marked restriction-code bound from Lemma~\ref{lem:punctured}, with
$
w+2R=56.
$
\end{enumerate}
The same certificate covers every first slice. The numbers of classes are given in Table~\ref{tab:quartic-certificates}.
\end{proposition}

\begin{table}[ht]
\centering
\caption{Complete covers for the exact-quartic derivative classes.}
\label{tab:quartic-certificates}
\begin{tabular}{c|c|c}
\hline
Classes & Certificate & Marked bound\\
\hline
12 & Two-erasure lemma & 55\\
68,268 & Ordinary background bound at most 54 & 55\\
151 & Point-dependent compensated restriction codes & 55\\
\hline
\end{tabular}
\end{table}

\begin{proof}
For the twelve analytic rows, the archived quadratic compensation has nonzero residual weight at most eighteen. Its zero set omits at least two points. Lemmas~\ref{lem:two-erasure} and~\ref{lem:punctured} give a background bound at most fifty-four, which increases by at most one after a point flip.

For 68,268 further rows, the complete numerical certificate verifies the ordinary assertion in the proposition. The remaining 151 rows use the marked assertions. Affine permutations preserving the derivative representative modulo quadratics act on its points. A verified subgroup suffices: each point is transported from one of 5,452 orbit representatives. Every generator, its quadratic offset, and all 19,328 point transporters are checked. The compensation may depend on the representative point.

For each restriction code the parity-check matrix is reconstructed from all cubic monomials. Independence and orthogonality give the exact kernel. The full syndrome domain is then covered by errors within the stated radius. Both arithmetic implementations pass all 73,720 selected tables; a fresh independent replay checks the same complete domains. Section~\ref{sec:verification} gives the encoding, arithmetic identities and reproducibility details. Classification completeness is the published input; it is not inferred from numerical row counts.
\end{proof}

\begin{theorem}
\label{thm:marked-quintic}
For every eight-variable background of degree at most five and every point,
$$
h\in\RM(5,8),\qquad p\in\F^8
\quad\Longrightarrow\quad
\nlr{3}(h+\dd_p)\le55.
$$
If the background has third-order nonlinearity fifty-six, then every point belongs to the support of a nearest cubic residual:
$$
\forall p\in\F^8\ \exists c\in\RM(3,8):
\quad\wt(h+c)=56,\quad(h+c)(p)=1.
$$
\end{theorem}

\begin{proof}
For exact degree five, use Lemma~\ref{lem:quartic-derivative} and the complete affine classification. The ordinary certificates in Proposition~\ref{prop:marked-certificate} give the marked bound by the triangle inequality; the marked certificates give it directly. All first slices and both positions of the marked point are retained.

For degree at most four, let a nearest cubic residual be nonzero. Its weight is at most fifty-six by Lemma~\ref{thm:external-quartic}. Translation changes a quartic by a cubic, so every translate is still a residual of the same background. Translate a support point onto the prescribed point and flip it. The weight decreases by one. A zero residual gives a singleton and also satisfies the assertion.

Finally, take a correction attaining the marked bound for a background at distance fifty-six. If its unmarked residual vanished at the prescribed point, its unmarked weight would be at most fifty-four. Therefore the residual contains the point, and its weight is exactly fifty-six. The triangle inequality also gives
$
\nlr{3}(h+\dd_p)=55.
$
\end{proof}

\begin{corollary}
\label{thm:r5}
Every function of degree at most five satisfies
$
\nlr{3}(h)\le56.
$
\end{corollary}

\begin{proof}
Apply Theorem~\ref{thm:marked-quintic} and change the prescribed point back. This argument does not use any seventh-order conclusion.
\end{proof}

\section{A strict local bound for exact-degree-six functions}
\label{sec:sextic}

The higher-degree relative argument needs a seven-variable estimate strictly smaller than the published cubic covering radius twenty. This local theorem remains a separate input: its domain has seven variables, whereas the relative theorem proved next has eight.

\begin{theorem}
\label{thm:sextic}
Every seven-variable Boolean function of exact degree six satisfies
$
\nlr{3}(g)\leq 18.
$
\end{theorem}

\begin{proof}
Use the complete affine classification of the quotient of all seven-variable Boolean functions by the cubic code~\cite{GillotLangevin2023}. The classification contains 3,486 affine classes, of which 1,711 have exact degree six. For every one of these 1,711 representatives, the certificate package contains an explicit cubic correction whose residual weight is at most eighteen. If an input is obtained from its class representative by an affine change of variables and addition of a cubic function, apply the same affine change to the stored correction and add the cubic function. The Reed--Muller code is invariant under affine changes, and Hamming weight is preserved. Therefore the transported correction remains cubic and has the same residual weight. This proves the bound for every function in every classified class.

The residual-weight histogram of the 1,711 certificates is recorded below. It documents the finite coverage, not exact distances.
$$
\begin{array}{c|rrrrrrrrr}
\text{residual weight}&2&4&6&8&10&12&14&16&18\\
\hline
\text{number of classes}&1&1&3&7&18&43&154&1316&168
\end{array}
$$
The first verification reconstructs each representative from its ANF, evaluates the correction on all 128 points, and checks its degree by a Boolean M\"obius transform. The second verification reconstructs the correction from its monomial support and independently repeats the truth-table and residual checks. Both verifications pass for all 1,711 classes. The completeness of the affine classification remains an external dependency.
\end{proof}

\section{Quadratic-label geometry and shared-slice compatibility}
\label{sec:r6}

The derivative-support argument of Section~\ref{sec:relative} controls targets of degree at most five. We now develop a geometric obstruction on quadratic spectral labels for the even-weight slicing argument. Its role is to improve a seven-variable quadratic approximation from forty to thirty-eight while retaining the cubic correction already chosen on the opposite slice.

\subsection{Odd six-variable functions and quadratic labels}

We use the following consequence of the complete six-variable cubic-coset classification and Table~1 of Gao, Kan, Li, and Wang~\cite{GaoKanLiWang2023}.

\begin{proposition}
\label{prop:odd-cosets}
For a six-variable function of odd weight, the possible third-order nonlinearities are
$$
1,\quad3,\quad5,\quad7.
$$
For each such cubic coset, the maximum second-order nonlinearity within the coset is, respectively,
$$
17,\quad15,\quad15,\quad15.
$$
In particular,
$$
\wt(g)\text{ odd},\quad\nlr{3}(g)>1
\quad\Longrightarrow\quad
\nlr{2}(g+c)\le15
\quad\text{for every }c\in\RM(3,6).
$$
\end{proposition}

Here maximization over homogeneous cubic corrections, as used in the published table, is equivalent to maximization over the whole cubic code: adding a quadratic does not change second-order nonlinearity. This is an external classification input, not a consequence of the spectral computations below.

\begin{lemma}
\label{lem:odd27}
An odd-weight six-variable function with first-order nonlinearity twenty-seven differs from a cubic bent function at one point. Consequently,
$$
\wt(g)\text{ odd},\quad\nlr{3}(g)>1
\quad\Longrightarrow\quad
\nlr{1}(g+q)\le25
\quad\text{for every }q\in\RM(2,6).
$$
\end{lemma}

\begin{proof}
Translate the input by its first moment, so that its first moment becomes zero. Write its Walsh coefficients as
$$
W(u)=\sum_{x\in\F^6}(-1)^{g(x)+u\cdot x}.
$$
For every linear hyperplane test, the parity of the restricted support is zero. The identity expressing a Walsh coefficient as the zero-frequency coefficient minus twice a restricted signed sum therefore gives
$
W(u)\equiv W(0)\pmod8.
$
First-order distance twenty-seven means that the largest absolute Walsh coefficient is ten. Taking the complement if necessary, all coefficients are congruent to two modulo eight. Their only possible values are
$$
-6,\quad2,\quad10.
$$
Let the respective multiplicities be three nonnegative integers. Parseval's identity and the sum of all Walsh coefficients give
$$
a+b+c=64,\qquad36a+4b+100c=4096,
\qquad-6a+2b+10c=64(-1)^{g(0)}.
$$
The unique nonnegative solution is
$$
g(0)=0,\qquad(a,b,c)=(36,0,28).
$$
Flipping the value at zero changes every Walsh coefficient by minus two, leaving only the values minus eight and eight. The resulting function is bent. To check that it is cubic, sum its signs over any affine four-flat. Walsh inversion expresses this sum as twice a sum of four signs, hence it is divisible by four. The function therefore has even weight on every affine four-flat, which is equivalent to degree at most three.

Finally, Parseval bounds the first-order distance of a six-variable function by twenty-eight. For odd-weight functions the distance is odd, so the only possible value above twenty-five is twenty-seven. Quadratic addition preserves the cubic coset. This proves the consequence.
\end{proof}

Let the quadratic-label space and the label cost of a six-variable function be
$$
\mathcal Q=\RM(2,6)/\RM(1,6),\qquad
D_g(q)=\nlr{1}(g+q).
$$
The cost is independent of the representative of a label. We use its two endpoint sets
$$
A_g=\{q\in\mathcal Q:D_g(q)=15\},\qquad
H_g=\{q\in\mathcal Q:D_g(q)=25\}.
$$
An affine change of input induces an invertible linear map on this quotient. Adding a quadratic translates both endpoint sets by the same label. Thus separate affine normalizations of two functions must retain the induced relative action on their label sets.

\subsection{A transport obstruction from common-point stars}

For a quadratic label, let its alternating polarization matrix and, in rank two, its image space be denoted by
$$
B_q,\qquad U(q)=\operatorname{im}B_q.
$$
For a subset of the label space, a root label and a nonzero point, define the common-point star by
$$
\mathcal S(S,a,p)=
\{a+b:b\in S,\ \rank B_{a+b}=2,\ p\in U(a+b)\}.
$$
The dimension of a star means the dimension of its linear span in the fifteen-dimensional label space, not the dimension of the image spaces.

\begin{lemma}
\label{lem:star-transport}
Suppose that an affine input permutation induces a label map and that a translated image of one label set is contained in another. Then every source star maps into the corresponding target star, and its span dimension is preserved by the map.
\end{lemma}

\begin{proof}
Let the linear part of the input permutation and its label map be
$$
L\in\operatorname{GL}(6,2),\qquad T:\mathcal Q\longrightarrow\mathcal Q.
$$
Polarization and its image transform according to
$$
B_{Tq}=L^{\mathsf T}B_qL,\qquad
U(Tq)=L^{\mathsf T}U(q).
$$
The affine translation contributes only an affine term to the transformed quadratic and hence disappears in the quotient. If
$
T(A)+v\subseteq H,
$
taking label differences cancels the label translation and gives
$$
T\mathcal S(A,a,p)
\subseteq\mathcal S(H,T(a)+v,L^{\mathsf T}p).
$$
The label map is invertible and linear, so it preserves span dimension. This proves the assertion for every legal input permutation without enumerating the linear group.
\end{proof}

We retain the numbering of the seven second-order-distance-fifteen representatives in Wang~\cite{Wang2019}, and denote them by
$
h_6,\ h_7,\ h_8,\ h_9,\ h_{10},\ h_{11},\ h_{12}.
$
Their explicit polynomials and the label encoding are given in Appendix~\ref{app:representatives}. Every six-variable function with second-order nonlinearity fifteen is affine equivalent, modulo a quadratic, to one of these representatives. Completeness of this list is the published classification input; our verification concerns the spectra and stars of the listed representatives.

\begin{proposition}
\label{prop:spectral-certificate}
The sixth representative in Wang's numbering is cubic plus a point indicator. The eighth has no label cost above twenty-three. For the remaining five representatives, Table~\ref{tab:stars} gives a source star dimension attained in the fifteen layer and the maximum star dimension over all roots and all nonzero points in the twenty-five layer.

\begin{table}[ht]
\centering
\caption{Complete endpoint and star data for the five relevant representatives.}
\label{tab:stars}
\begin{tabular}{lrrrr}
\hline
Type & Fifteen labels & Twenty-five labels & Source span & Target span\\
\hline
Seventh &96&1024&5&4\\
Ninth &72&512&5&3\\
Tenth &72&256&5&3\\
Eleventh &40&544&5&3\\
Twelfth &66&414&5&3\\
\hline
\end{tabular}
\end{table}

Consequently, the fifteen layer of any of these five representatives cannot be contained, after any legal label transport and translation, in the twenty-five layer of any of the five representatives.
\end{proposition}

\begin{proof}
The finite assertions are verified over the complete domains specified in Section~\ref{subsec:spectral-verification}; the source witnesses are listed in Appendix~\ref{app:representatives}. The containment assertion follows from Lemma~\ref{lem:star-transport}: a five-dimensional source star cannot map into a target star of dimension at most four. Importantly, the source and target representatives need not have the same type or the same affine normalization.
\end{proof}

\subsection{The local improvement from forty to thirty-eight}

\begin{theorem}
\label{thm:odd38}
If two odd-weight six-variable functions are not in a cubic-plus-singleton coset, their concatenation has second-order nonlinearity at most thirty-eight:
$$
\nlr{3}(g_0)>1,\quad\nlr{3}(g_1)>1
\quad\Longrightarrow\quad
\nlr{2}(g_0\Vert g_1)\le38.
$$
\end{theorem}

\begin{proof}
The concatenation is a seven-variable function. A quadratic correction has a common homogeneous quadratic label on the two slices and independent affine tails. Therefore
$$
\nlr{2}(g_0\Vert g_1)
=\min_{q\in\mathcal Q}\{D_{g_0}(q)+D_{g_1}(q)\}.
$$
Proposition~\ref{prop:odd-cosets} bounds each minimum label cost by fifteen, and Lemma~\ref{lem:odd27} bounds every label cost by twenty-five. If either minimum is at most thirteen, the displayed identity gives thirty-eight immediately. The minima are odd, so otherwise both are fifteen.

Apply the published seven-type classification to each slice separately. The sixth representative is excluded by the hypothesis on the cubic coset. If either slice has the eighth type, its costs are all at most twenty-three, and a fifteen label of the other slice gives the result.

It remains to consider two of the five types in Table~\ref{tab:stars}. If every fifteen label of the first slice had cost twenty-five on the second, there would be a containment
$
A_{g_0}\subseteq H_{g_1}.
$
After separate affine normalizations, this is precisely a transported and translated containment forbidden by Proposition~\ref{prop:spectral-certificate}. Hence some fifteen label has cost at most twenty-three on the other slice. Using that common label and its independently optimal affine tails gives
$
\nlr{2}(g_0\Vert g_1)\le15+23=38.
$
\end{proof}

This theorem isolates the compatibility gain. It does not add two independently optimized quadratic approximations: it finds a shared quadratic label. The point-star invariant supplies the obstruction that forces such a label, while the finite certificate establishes its hypotheses for the complete list of endpoint types.

\section{Moment-controlled slices and singleton rigidity}
\label{sec:unified}

We prove a common upper-bound argument for even-weight eight-variable inputs. It uses the local theorems above without presupposing the sixth-order relative theorem.

\subsection{Moments and paired normal planes}

Define binary parity, first moment and second moment by
$$
p(F)=\sum_x F(x),\qquad
v=m(F)=\sum_x xF(x),\qquad
M=M(F)=\sum_x xx^{\mathsf T}F(x).
$$
Cube summation of the algebraic normal form gives
$$
p(F)=\alpha_{[8]},\qquad
m_j(F)=\alpha_{[8]}+\alpha_{[8]\setminus\{j\}}.
$$
Even parity and zero first moment characterize membership in the sixth-order code. When these moments vanish, the off-diagonal second moments are the complementary sextic coefficients. Vanishing of all three moments therefore characterizes membership in the fifth-order code.

Assume throughout this subsection that parity is zero. Define the normal space and its restricted form by
$$
H=v^\perp,\qquad
A(\alpha,\beta)=\alpha^{\mathsf T}M\beta
\quad(\alpha,\beta\in H).
$$
The diagonal of the second moment is the first moment, so the restricted form is alternating. Translation by an input vector changes the second moment according to
$$
M\longmapsto M+va^{\mathsf T}+av^{\mathsf T},
$$
and hence leaves its restriction to the normal space unchanged. Linear input changes transport that form by invertible congruence.

\begin{lemma}
\label{lem:paired-slices}
If two normals in the normal space pair to one, their four parallel six-dimensional restrictions all have odd weight. Each containing seven-dimensional hyperplane is an exact sextic.
\end{lemma}

\begin{proof}
The hyperplane indicators give
$$
\sum_{\alpha\cdot x=b}F(x)=(1+b)p(F)+\alpha\cdot v=0,
$$
$$
\sum_{\alpha\cdot x=b,\ \beta\cdot x=c}F(x)
=\alpha^{\mathsf T}M\beta=1.
$$
The parity and first-moment contributions vanish for every pair of offsets. Each containing seven-dimensional restriction is even and has degree at most six. Its odd six-dimensional restriction prevents its degree from being at most five.
\end{proof}

\begin{lemma}
\label{lem:nonsingleton-alternative}
For a paired normal plane, if at least one of its four odd six-dimensional restrictions is outside a cubic-plus-singleton coset, the original input has third-order nonlinearity at most fifty-six.
\end{lemma}

\begin{proof}
If all four are outside these cosets, pair them into two seven-dimensional hyperplanes. Correct one cubically with cost at most eighteen by Theorem~\ref{thm:sextic}, and extend that cubic constantly to the other. Its two six-dimensional restrictions remain nonsingleton. Theorem~\ref{thm:odd38} supplies a quadratic correction of cost at most thirty-eight on the other hyperplane. With the first hyperplane at outer coordinate zero, the global correction has the legal form
$$
C(y,t)=a(y)+tq(y),\qquad \deg a\le3,\quad\deg q\le2.
$$
Its residual weight is at most
$
18+38=56.
$
Otherwise there is a singleton slice and a nonsingleton slice. Any two distinct cells of the two-dimensional index space can be paired by a linear change of index coordinates. Pair these two and use the singleton slice's cubic background on both. The singleton residual costs one. Proposition~\ref{prop:odd-cosets} gives a quadratic correction of cost at most fifteen on its nonsingleton partner after that common cubic addition. Multiplying this quadratic by the inner selector gives a cubic on the union. Extend it to the opposite seven-dimensional hyperplane and correct the latter quadratically with cost at most forty, using Lemma~\ref{thm:external-radii}. The outer selector again raises the degree only to three. The total is
$
1+15+40=56.
$
\end{proof}

\subsection{Graph rigidity and consistent singletons}

\begin{lemma}
\label{lem:graph-rigidity}
Use eight coordinates arranged as five base coordinates, one free coordinate and two selector coordinates. Suppose a polynomial of degree at most five restricts to a cubic on every six-dimensional graph obtained by making each selector an affine linear function of the five base coordinates, independently of the free coordinate. Then the polynomial is globally cubic.
\end{lemma}

\begin{proof}
Use the squarefree decomposition
$$
(z_1,\ldots,z_5,t,s_1,s_2),\qquad
K=K_0(z,t)+s_1B(z,t)+s_2C(z,t)+s_1s_2D(z,t).
$$
The zero-selector graph and the total degree give
$$
\deg K_0\le3,\qquad \deg B,\deg C\le4,\qquad \deg D\le3.
$$
Set the second selector to zero and the first to each base coordinate. Subtracting the cubic base restriction gives
$$
\deg(z_iB)\le3\qquad(1\le i\le5).
$$
Every degree-four monomial of the first coefficient polynomial misses at least one base coordinate. Multiplication by a missing coordinate produces a unique degree-five term: monomials containing the multiplier produce degree at most four. Thus all degree-four coefficients vanish. Repetition at degree three eliminates all degree-three coefficients. Interchanging selectors gives the same conclusion for the second coefficient polynomial. Hence
$$
\deg B\le2,\qquad \deg C\le2.
$$
Their selector products are globally cubic and may be subtracted. Substituting two distinct base coordinates for the selectors now gives
$$
\deg(z_i z_jD)\le3\qquad(i\ne j).
$$
Every degree-three monomial of the last coefficient polynomial misses at least two base coordinates. Multiplication by those two creates a unique degree-five term, so all degree-three coefficients vanish. Repetition at degree two eliminates all degree-two coefficients. Therefore
$$
\deg D\le1,\qquad \deg K\le3.
$$
Terms involving the free coordinate use fewer base coordinates and obey the same argument. The proof needs only thirty-six zero-offset graph choices: each selector is either zero or one of the five base coordinates.
\end{proof}

\begin{lemma}
\label{lem:singleton-graphs}
In the coordinates of Lemma~\ref{lem:graph-rigidity}, suppose every graph six-flat restricts the input to a cubic plus a singleton. Then the input is a global cubic plus the indicator of four points.
\end{lemma}

\begin{proof}
The flats have the form
$$
P_{a,b;c,d}=\{s_1=a(z)+c,\ s_2=b(z)+d\},
$$
where the two slopes are arbitrary linear forms and the free coordinate is unrestricted. There are one thousand twenty-four slope pairs. Each singleton location is unique, since a nonzero cubic on six variables has weight at least eight.

Fix an ambient point and compare two graph flats through it with either first slopes or second slopes equal. Distinct such flats intersect in dimension five. Their two cubic-plus-singleton descriptions, restricted to the intersection, differ by a cubic supported on at most two points. A nonzero cubic on five variables has weight at least four. The restricted singleton indicators therefore agree. In particular, a point marked on one flat is marked on the other.

The slope-pair graph is connected: change one slope and then the other. Thus whether the point is marked is independent of the chosen flat through it. Denote the globally marked set by $S_*$. Every graph flat contains exactly one globally marked point. The four zero-slope flats partition the space, so
$$
|S_*|=4.
$$
Remove these four points. The remainder is cubic on every graph flat. Interpolation on the zero-slope partition first gives total degree at most five, and Lemma~\ref{lem:graph-rigidity} makes it globally cubic.
\end{proof}

\begin{theorem}
\label{thm:nonzero-moment}
An even-weight eight-variable function with nonzero restricted second-moment form has third-order nonlinearity at most fifty-six.
\end{theorem}

\begin{proof}
Suppose its distance is greater than fifty-six. Lemma~\ref{lem:nonsingleton-alternative} forces all four restrictions on every paired normal plane to be cubic plus singleton. Choose one plane and interpolate the four cubic backgrounds:
$$
F=Q+\boldsymbol1_S,\qquad \deg Q\le5,\qquad |S|=4.
$$
The background has zero parity, first moment and second moment. Thus the moment data of the four points agree with those of the input. Translate one point to zero and write
$$
S=\{0,a,b,c\},\qquad v=a+b+c.
$$
On the normal space, the functional of the third vector is the sum of those of the first two. Consequently
$$
A(\alpha,\beta)
=(\alpha\cdot a)(\beta\cdot b)
+(\alpha\cdot b)(\beta\cdot a).
$$
Its rank is at most two, and nonvanishing makes it exactly two. This works for both zero and nonzero first moment.

Choose a seven-dimensional subspace of the normal space consisting of a symplectic pair and five radical directions. For nonzero first moment this is the full normal space; for zero first moment discard one of the six radical directions. Complete these seven normals to eight dual coordinates as in Lemma~\ref{lem:graph-rigidity}, with selector normals paired to one and base normals in the radical.

Every graph flat of Lemma~\ref{lem:singleton-graphs} is then defined by a paired normal plane. All its restrictions are cubic plus singleton. That lemma makes the entire input a cubic plus four points, giving
$
\nlr{3}(F)\le4,
$
a contradiction.
\end{proof}

\subsection{The zero restricted-moment branch}

A parity-sensitive extension of the punctured-code argument handles this branch. The complete cubic-mod-quadratic classification has twelve seven-variable representatives~\cite{GillotLangevin2023}. Ten require finite even-target checks; the other two are zero and a single cubic monomial modulo quadratics. The following class numbers are zero-based indices in the archived classification file.

\begin{proposition}
\label{prop:cubic-even-cover}
For every row of Table~\ref{tab:cubic-even}, the supplement supplies a cubic representative and a quadratic compensation. Their sum has the displayed weight. On its zero set, every even target is within the displayed radius of the punctured cubic code. The final column is the minimum unscaled ball count over all even target syndromes.

\begin{table}[ht]
\centering
\caption{Complete even-target covers for cubic derivatives.}
\label{tab:cubic-even}
\begin{tabular}{rrrrr}
\hline
Class & Weight & Radius & Dual dimension & Minimum count\\
\hline
0&44&6&21&40\\
1&44&6&21&26\\
2&44&6&21&24\\
3&44&6&21&24\\
4&40&8&25&832\\
5&44&6&21&30\\
6&40&8&26&192\\
7&40&8&25&960\\
8&40&8&26&128\\
9&40&8&26&192\\
\hline
\end{tabular}
\end{table}
\end{proposition}

\begin{proof}
Reconstruct the punctured cubic generator and the full parity-check kernel by rank and orthogonality. The exact syndrome ball-count identity of Section~\ref{subsec:syndrome-arithmetic} gives the positive even minima. The ten rows cover 278,921,216 syndromes, including odd syndromes, whose counts need not all be positive. Section~\ref{subsec:cubic-verification} specifies the checks and parity convention. These finite assertions do not reprove completeness of the twelve-type classification.
\end{proof}

\begin{lemma}
\label{lem:cubic-odd-first}
If an eight-variable function has an odd-weight seven-variable first slice and slice difference of degree at most three, its third-order nonlinearity is at most fifty-six.
\end{lemma}

\begin{proof}
Apply the complete cubic-mod-quadratic classification simultaneously to the slices, absorbing its quadratic ambiguity into a legal cubic correction. For a finite row write
$$
F(y,t)=f(y)+t g(y),\qquad q=g+b,\qquad \deg b\le2,\qquad
Z=\{y:q(y)=0\}.
$$
The exact residual identity for any cubic first-slice correction is
$$
\wt\bigl(F+a+t b\bigr)=\wt(q)+2\wt_Z(f+a).
$$
The seven-variable cubic code is self-dual: products of cubics have degree at most six and its dimension is sixty-four. The zero-set indicator is cubic, so every punctured cubic codeword has even weight. Restricted target parity is therefore a punctured-coset invariant.

Choose a nonzero cubic monomial of the residual. Its three coordinate directions span a translation space satisfying the third-difference identity
$$
\sum_{h\in U}q(y+h)=1,\qquad |U|=8.
$$
Each translated residual has the same even-target certificate and is a legal compensation, since
$$
b_h(y)=g(y)+q(y+h)\in\RM(2,7).
$$
If the odd first slice were odd on every translated zero set, the sum of these eight parities would be zero. But it is also
$$
\sum_{h\in U}\sum_y f(y)\bigl(1+q(y+h)\bigr)
=\sum_y f(y)=1,
$$
a contradiction. At least one translated zero set has an even target restriction. Proposition~\ref{prop:cubic-even-cover} and the residual identity give
$
\wt(q)+2R=56.
$
For the single-monomial type, use zero compensation. The residual has weight sixteen; puncturing cannot increase the full cubic radius twenty. Hence
$
16+2\cdot20=56.
$
For the zero type, absorb the quadratic derivative entirely. The two equal odd slices give
$
2\cdot19=38.
$
Every affine map and translation is applied simultaneously to the slices, retaining all lower coefficients and relative positions.
\end{proof}

\begin{lemma}
\label{lem:zero-moment}
An even-weight eight-variable function with zero restricted second-moment form has third-order nonlinearity at most fifty-six.
\end{lemma}

\begin{proof}
If the first moment vanishes, the normal space is the whole dual space. The second moment is zero, so the degree is at most five and Corollary~\ref{thm:r5} applies.

Otherwise take the last coordinate direction along the nonzero first moment. The coefficient identities give
$$
F(y,t)=f(y)+t g(y),\qquad \wt(f)\text{ odd},\qquad \deg g\le5.
$$
For quotient-coordinate indices the restricted second moments are
$$
M_{ij}=\sum_y y_i y_j g(y).
$$
These are exactly the complementary quintic coefficients of the derivative. Their vanishing forces derivative degree at most four. For exact quartic degree, Section~\ref{subsec:relative-certificates} and Lemma~\ref{lem:punctured} give a correction uniform in the arbitrary first slice. For degree at most three, apply Lemma~\ref{lem:cubic-odd-first}.
\end{proof}

\subsection{The even-weight upper bound}

\begin{theorem}
\label{thm:r7}
Every even-weight eight-variable function satisfies
$
\nlr{3}(F)\le56.
$
\end{theorem}

\begin{proof}
Even weight is equivalent to degree at most seven. Apply Theorem~\ref{thm:nonzero-moment} or Lemma~\ref{lem:zero-moment}, according as the second-moment form restricted to the first-moment annihilator is nonzero or zero.
\end{proof}

\section{Odd functions and centered covariance}
\label{sec:odd-moments}

The preceding even-weight bound leaves only a possible odd distance fifty-seven. We show that nonzero centered second moment excludes this value. The zero-moment case then falls under Theorem~\ref{thm:marked-quintic}.

\subsection{The common-cubic quotient}

For an odd seven-variable function, define its first moment and centered covariance by
$$
v(O)=\sum_x xO(x),\qquad
\Omega(O)=\sum_x xx^{\mathsf T}O(x)+v(O)v(O)^{\mathsf T}.
$$
These sums are binary. The covariance is alternating and is invariant under translation; linear changes act by congruence. Cubic addition changes none of these moments on seven variables.

\begin{proposition}
\label{prop:odd-covariance}
For an odd seven-variable function,
$$
\nlr{3}(O)=19,\qquad\Omega(O)\ne0
\quad\Longrightarrow\quad
\nlr{2}(O+c)\le37
\quad\text{for every }c\in\RM(3,7).
$$
\end{proposition}

\begin{proof}
The complete classification of
$$
\RM(7,7)/\RM(3,7)
$$
contains 3,486 affine classes~\cite{GillotLangevin2023}. Explicit cubic corrections give upper bound seventeen for all but eleven of the odd classes. Of these eleven necessary endpoint classes, eight have nonzero centered covariance. Their one-based catalogue indices are
$$
1,\ 1000,\ 1303,\ 1304,\ 1355,\ 1362,\ 1538,\ 1547.
$$
For each of the eight representatives we verify every cubic addition modulo quadratics. This is a thirty-five-dimensional quotient:
$$
V=\RM(3,7)/\RM(2,7).
$$
Write the representative in two six-variable slices. Let the common homogeneous cubic, the relative homogeneous quadratic, and the shared quadratic label be denoted by
$$
P\in\mathcal H_3(6),\qquad
s\in\mathcal H_2(6),\qquad q\in\mathcal H_2(6),
$$
where the homogeneous spaces have dimensions twenty and fifteen. Define exact affine-distance profiles by
$$
A_P(q)=\nlr{1}(f_0+P+q),\qquad
B_P(q)=\nlr{1}(f_1+P+q).
$$
A seven-variable quadratic has a common homogeneous quadratic and independent affine tails on its two slices. Consequently the exact formula is
$$
\nlr{2}(O+P+t s)
=\min_q\bigl(A_P(q)+B_P(q+s)\bigr).
$$
The certificate verifies that this minimum is at most thirty-seven for every relative label and every common core. Verified affine subgroups cover the common-core space; every generator acts on the full thirty-five-dimensional quotient and carries the entire relative-label fibre bijectively. Full stabilizer completeness is unnecessary. Each certified common-core fibre retains every relative quadratic label. The independent subgroup and arithmetic checks cover the whole coefficient domain, as specified in Section~\ref{sec:verification}. Scalar certificates use Lemma~\ref{lem:odd27} and the six-variable bent degree bound~\cite{Carlet2021}; the other certificates use exact profile convolutions or explicit covering sets. The input associations, subgroup parent edges, closure edges and numerical records are checked separately. Appendix~\ref{app:covariance-inputs} specifies the representatives, and Section~\ref{sec:verification} specifies the finite obligations.

Finally, affine changes preserve quadratic distance, transport the whole cubic code, and preserve the nonvanishing of covariance. Addition of a cubic translates the complete cubic-label space. Thus the verified representative assertion gives the stated implication for every function in each classified coset and every affine position.
\end{proof}

\subsection{Endpoint saturation on a normal plane}

\begin{lemma}
\label{lem:seven-singletons}
Let an even seven-variable function have nonzero first moment. If all its odd six-dimensional hyperplane restrictions are cubic plus a singleton, then the function is cubic plus two points.
\end{lemma}

\begin{proof}
Choose the first-moment direction as the last coordinate. The odd hyperplanes are exactly the graphs
$$
s=a(y)+c,\qquad y\in\F^6,
$$
with a linear function and either constant. On each graph the singleton is unique, since a nonzero cubic on six variables has weight at least eight. Two distinct graph directions through a point meet in a five-flat. Their singleton indicators must agree there: otherwise their difference would be a cubic of weight one or two, smaller than four. Any graph mark agrees with the mark of the constant-selector hyperplane through it. Conversely, a constant-selector mark lying on another graph must be that graph's mark. The two global marks lie one in each constant-selector hyperplane, and every graph contains exactly one of them.

Remove these two points. The remainder is cubic on every graph. Its two constant-selector restrictions give
$$
K(y,s)=K_0(y)+sB(y),\qquad\deg(K_0)\le3,\quad\deg(B)\le3.
$$
Each substitution of a base coordinate for the selector is cubic. Any cubic monomial in the selector coefficient misses a base coordinate; multiplying by that coordinate creates a unique quartic coefficient. Hence all cubic coefficients of that selector coefficient vanish. The remainder is globally cubic.
\end{proof}

For an odd eight-variable function, translate its first moment to zero. Define
$$
p(F)=1,\qquad m(F)=0,\qquad
M=\sum_x xx^{\mathsf T}F(x),\qquad
T_\alpha=\sum_x(\alpha\cdot x)xx^{\mathsf T}F(x).
$$
The second moment is alternating. For a normal outside its radical, let the through-zero and opposite restrictions be denoted by
$$
O_\alpha=F|_{\alpha\cdot x=0},\qquad
E_\alpha=F|_{\alpha\cdot x=1}.
$$
The first is odd. The second is even with nonzero first moment, hence is an exact sextic.

\begin{lemma}
\label{lem:odd-endpoints}
If the original function has third-order nonlinearity fifty-seven, every such pair satisfies
$$
\nlr{3}(O_\alpha)=19,\qquad\nlr{3}(E_\alpha)=18.
$$
\end{lemma}

\begin{proof}
The upper bounds follow from parity and Theorem~\ref{thm:sextic}. An even-side upper bound sixteen would combine with the odd-side quadratic upper bound thirty-nine, using the same cubic and a selector times a quadratic, to give fifty-five. Thus the even endpoint is eighteen.

Choose a partner normal pairing to one with the fixed normal. Its four six-dimensional cells have one even and three odd parities, since
$$
p_{b,c}=(1+b)(1+c)+\alpha^{\mathsf T}M\beta.
$$
If the three odd cells contain both a singleton and a nonsingleton cubic coset, pair these two into an even seven-hyperplane. The compatible cost is at most one plus fifteen. The other seven-hyperplane is odd and admits quadratic cost at most thirty-nine after the same cubic. The total is at most fifty-five.

Suppose now that the odd seven-side had cubic cost at most seventeen. For any partner plane whose three odd cells are nonsingleton, the even seven-side has uniform quadratic cost at most thirty-eight by Theorem~\ref{thm:odd38}. The common correction again gives fifty-five. Therefore all three odd cells must be singletons for every partner plane. The odd hyperplanes of the even restriction are precisely those obtained by these partner normals. Lemma~\ref{lem:seven-singletons} would make that restriction cubic plus two points, contradicting its cubic distance eighteen. The odd endpoint must consequently be nineteen.
\end{proof}

\subsection{A covariance-rank obstruction}

The odd through-zero restriction has first moment and centered covariance, embedded in the ambient coordinates,
$$
v_\alpha=M\alpha,\qquad
\Omega_\alpha=M+T_\alpha+v_\alpha v_\alpha^{\mathsf T}.
$$
Symmetry of the third moment gives
$
T_\alpha\beta=T_\beta\alpha.
$

\begin{lemma}
\label{lem:covariance-rank}
If the centered second moment has positive rank, the number of nonradical normals whose odd restriction has zero centered covariance is at most
$$
(r+1)2^{8-r},\qquad r=\rank(M).
$$
In particular, at ranks four, six and eight there is a nonradical normal with nonzero odd-side covariance.
\end{lemma}

\begin{proof}
Suppose two zero-covariance normals have distinct first-moment images. Put
$$
v=M\alpha,\qquad w=M\beta,\qquad s=\alpha^{\mathsf T}M\beta.
$$
The covariance equations and tensor symmetry imply
$$
w+sv=v+sw,\qquad (1+s)(v+w)=0.
$$
Thus their pairing is one. Choose one normal for every distinct image occurring in the zero-covariance set. If their number is even, their Gram matrix, with zero diagonal and all other entries one, has full rank; if it is odd, its rank is one less. Since this rank cannot exceed the even rank of the ambient form, the number of images is at most one more than that rank. Every image has at most the kernel size as its number of preimages. This proves the bound. For ranks four, six and eight it is strictly smaller than the number of nonradical normals.
\end{proof}

\begin{lemma}
\label{lem:rank-two-odd}
Suppose the centered second moment has rank two and all its nonradical odd restrictions have zero covariance. Then
$$
F=Q+\boldsymbol1_{A\setminus\{0\}},\qquad
\deg(Q)\le4,\qquad\dim(A)=2,
$$
where the two-plane is linear and contains the centered point.
\end{lemma}

\begin{proof}
For a radical normal, add it to a fixed nonradical normal. The first-moment images are equal, so subtracting their covariance equations gives zero contraction of the third moment along every radical normal. Normalize the alternating form so that its only nonzero pair is in the first two coordinates. Tensor symmetry then makes every third moment on three distinct coordinates zero.

Remove the centered singleton. Parity and first moment show that the remainder has degree at most six. The second moments give its sextic part, and the vanishing third moments give its quintic part:
$$
P_6=x_3x_4x_5x_6x_7x_8,\qquad
P_5=\sum_{k=3}^8\prod_{\substack{3\le j\le8\\j\ne k}}x_j.
$$
For the latter assertion, summing a monomial against three distinct coordinates selects the complementary quintic coefficient, together with the unique sextic coefficient when the triple contains both first coordinates. These are exactly the degree-six and degree-five parts of the plane indicator
$$
\boldsymbol1_A=\prod_{k=3}^8(1+x_k),\qquad A=\langle e_1,e_2\rangle.
$$
After subtracting this indicator, only degree at most four remains. Adding back the centered singleton proves the assertion.
\end{proof}

\begin{lemma}
\label{lem:quartic-three-points}
Every quartic background plus three points of an affine two-plane has third-order nonlinearity at most fifty-five.
\end{lemma}

\begin{proof}
Translate the removed point of the plane to zero. Choose a nearest cubic residual of the quartic background, of weight at most fifty-six. Every translate is a residual of the same background, since the translation difference of a quartic is cubic. If its weight is zero, the three-point residual suffices. If its positive weight is at most fifty-four, translate one support point into the three-point set, obtaining total weight at most fifty-five.

Suppose the residual has weight fifty-six. Some plane coset must contain at least two support points. Otherwise each plane coset contains at most one, and summation over the plane is also the ordinary count of support points in that coset. It defines a quadratic on the six-dimensional quotient with weight fifty-six:
$$
G(z)=D_uD_vr(z),\qquad\deg(G)\le2,\qquad\wt(G)=56.
$$
A quadratic on six variables has signed sum zero or a signed power of two: squaring its signed sum and summing its linear derivatives over the radical of the polar form proves this immediately. Weight fifty-six would instead give signed sum minus forty-eight, a contradiction. Two support points therefore lie in one plane coset. Every nonzero plane direction occurs as a difference of two points in its punctured three-point set. Translate the support pair onto such a pair. The final weight is at most
$
56+3-2\cdot2=55.
$
\end{proof}

\begin{theorem}
\label{thm:odd-nonzero-moment}
An odd eight-variable function with nonzero centered second moment satisfies
$
\nlr{3}(F)\le55.
$
\end{theorem}

\begin{proof}
The even-weight theorem and a singleton change give an odd upper bound fifty-seven. Suppose equality holds. Lemma~\ref{lem:odd-endpoints} gives the coupled endpoints for every nonradical normal. If any odd restriction has nonzero covariance, apply Proposition~\ref{prop:odd-covariance} after choosing an eighteen correction on the actual even partner. The same cubic and an additional odd-side quadratic give the compatible cost
$
18+37=55.
$
At second-moment ranks four, six and eight, Lemma~\ref{lem:covariance-rank} supplies such a normal. At rank two, either it exists or Lemma~\ref{lem:rank-two-odd} gives the quartic-plus-three-point form, which is covered by Lemma~\ref{lem:quartic-three-points}. These exhaust the nonzero alternating ranks. Equality at fifty-seven is impossible; odd parity gives the claimed bound.
\end{proof}

\section{Proof of the covering-radius theorem}
\label{sec:unrestricted}

\begin{proof}[Proof of Theorem~\ref{thm:global}]
Even-weight inputs have upper bound fifty-six by Theorem~\ref{thm:r7}. Let the input have odd weight. Translation changes its first moment by the translation vector, so we may center that moment at zero. If the centered second moment is nonzero, Theorem~\ref{thm:odd-nonzero-moment} gives fifty-five.

Otherwise remove the centered singleton. The resulting even function has zero first and second moments. Cube summation of its ANF therefore puts it in the fifth-order code:
$$
F=h+\dd_0,\qquad h\in\RM(5,8).
$$
Theorem~\ref{thm:marked-quintic} gives fifty-five in this case as well. Thus every odd input has third-order nonlinearity at most fifty-five, and every input has upper bound fifty-six. The matching lower bound is the published quartic relative value in Lemma~\ref{thm:external-quartic}.

Finally, exact degree eight is equivalent to odd truth-table weight. Choose a quartic function at distance fifty-six and flip any point. The new function has exact degree eight, and the triangle inequality gives distance at least fifty-five. The odd upper bound proves
$$
\rho\bigl(\RM(3,8)\bigr)=56,
\max_{\deg F=8}\nlr{3}(F)=55.
$$
\end{proof}

\section{Finite verification and reproducibility}
\label{sec:verification}

The proof uses complete finite assertions at four local interfaces: marked quartic-derivative covers, strict seven-variable sextic corrections, six-variable quadratic-label stars, and the seven-variable nonzero-covariance endpoint. The moment reductions, rank obstruction, singleton consistency and gluing of corrections are proved in the text. The finite assertions and the published classifications have different roles.

\subsection{Classification inputs and complete domains}

Gillot and Langevin~\cite{GillotLangevin2023} supply the complete affine classifications of the quotients
$$
\RM(4,7)/\RM(2,7),\qquad
\RM(7,7)/\RM(3,7),\qquad
\RM(3,7)/\RM(2,7).
$$
Their numbers of classes are, respectively,
$$
68,443,\qquad3,486,\qquad12.
$$
These are classification theorems used as external completeness inputs. The first quotient has dimension seventy; its 68,431 exact-quartic classes form the derivative domain in Proposition~\ref{prop:marked-certificate}. The second has dimension sixty-four. It supplies all 1,711 exact-sextic classes for Theorem~\ref{thm:sextic}, and the complete odd-class list used in Proposition~\ref{prop:odd-covariance}. The third has dimension thirty-five and supplies the cubic-derivative classes in Lemma~\ref{lem:cubic-odd-first}. It is also the space of relative cubic labels in the covariance endpoint; its twelve affine classes cannot replace the relative-label verification for a fixed odd function.

The seven six-variable representatives of second-order nonlinearity fifteen and the odd cubic-coset maxima are published inputs~\cite{Wang2019,GaoKanLiWang2023}. Their use in the star argument is specified in Section~\ref{sec:r6}. No classification of the full eight-variable quotient is assumed.

\subsection{Exact restriction-code certificates}
\label{subsec:syndrome-arithmetic}

For each selected table, form the evaluation matrix of all cubic monomials on its specified coordinates. A parity-check matrix is accepted only after independence, orthogonality and complementary dimension establish its exact kernel. Covering at a stated radius is then equivalent to the full syndrome identity
$$
B_R=\F^d,\qquad
B_R=\{He:\wt(e)\le R\}.
$$
Two exact arithmetic forms are used. The integer-count implementation inserts the syndrome indicator by character orthogonality. With the punctured length and Krawtchouk polynomials, the exact formula is
$$
2^dN_R(s)=\sum_{z\in\F^d}(-1)^{z\cdot s}
\sum_{j=0}^R K_j\bigl(\wt(H^{\mathsf T}z)\bigr),
$$
$$
K_j(w)=\sum_{\ell=0}^j(-1)^\ell
\binom{w}{\ell}\binom{n-w}{j-\ell},\qquad n=|S|.
$$
Indeed, substitute the binary character expansion of the indicator of the syndrome equation and sum over every error of weight at most the radius. Positivity for every syndrome proves the covering assertion. Integer arithmetic checks nonnegativity and exact divisibility; intermediate bounds are checked before a signed sixty-four-bit transform is used. The support implementation uses the identity
$
B_{a+b}=B_a+B_b.
$
To prove it, split an error support into two parts for one inclusion. For the other, the XOR of two errors has weight at most the sum of their weights, even if their supports overlap. Thus overlap does not invalidate the decomposition. Complete small-weight syndrome sets and exact bitset shifts verify their union. Early termination is permitted only when no syndrome is uncovered. This establishes positivity of every syndrome count, not its exact minimum multiplicity.

The marked-derivative certificate has 68,268 ordinary background rows and 151 marked rows. The latter use 5,452 point-orbit representatives. Both implementations verify 73,720 selected tables, hence 147,440 tables in total. Per arithmetic path their complete syndrome domains have total size
$$
747,008,163,840.
$$
The largest dual dimension is twenty-six. Each quadratic compensation is checked against the original derivative ANF. Every affine subgroup generator is checked by direct substitution modulo quadratics, and every point has a saved generator-word transporter. The first slice is never fixed or sampled. Twelve further rows use Lemma~\ref{lem:two-erasure}, rather than a numerical covering assertion.

A complete fresh independent replay recomputes all 73,720 selected tables, with previously completed archives reused only after their input keys, compressed hashes, raw hashes and dual bases have been checked. The final archive association check again matches every table to its exact input, point, radius and dual basis. The independent implementation changes the pivot order, combination generation and the way uncovered syndromes are filtered. The supplement records which outputs are exact multiplicities and which are exact support certificates.

\subsection{Sextic corrections and quadratic-label stars}
\label{subsec:spectral-verification}

All 1,711 seven-variable sextic inputs are reconstructed from their ANFs. Each supplied cubic correction is evaluated on all 128 points and independently checked by a Boolean M\"obius transform. These are upper-eighteen certificates; they do not assert exact distance eighteen for each representative.

The six-variable certificate reconstructs the seven representatives in Appendix~\ref{app:representatives}, all quadratic labels and their Walsh spectra. The star verification checks every target root and every nonzero point, records full target-star spans, and compares them with explicit five-dimensional source stars. Lemma~\ref{lem:star-transport} then carries the finite obstruction to every relative affine position. A histogram match would not provide that transport.

\subsection{Parity-sensitive cubic covers}
\label{subsec:cubic-verification}

For the twelve cubic-derivative classes, ten finite rows and two analytic rows establish the parity-sensitive covering statement of Lemma~\ref{lem:cubic-odd-first}. Every supplied dual basis and each of the eight translated compensations are checked. The ten finite rows cover 278,921,216 syndromes per complete pass. The supplementary directories \texttt{r6/} and \texttt{r7/} retain these certificates and their independent checkers.

\subsection{The covariance endpoint certificate}

For the odd seven-variable catalogue, explicit cubic corrections give upper seventeen outside the eleven necessary endpoint classes. Covariance removes three of these from the hypothesis of Proposition~\ref{prop:odd-covariance}; no nineteen lower bound for every remaining representative is required.

For each of the eight applicable representatives, the full relative-label space decomposes as
$$
\mathcal H_3(7)=\mathcal H_3(6)\oplus t\mathcal H_2(6).
$$
All lower quadratic coefficients are retained by exact affine minimization. Verified affine subgroups act on the whole thirty-five-dimensional quotient, with a common-core action independent of the relative quadratic label. For a fixed common core the latter is transported bijectively, so a whole fibre may be carried together. The certificate checks every generator on the constant and all basis directions, every parent edge, every closure edge and every required core orbit. A subgroup cover suffices without a full stabilizer computation.

For a fixed common core, the exact number of accepted shared labels for a relative label is
$$
\begin{aligned}
C_P(s)&=\sum_q\boldsymbol1_{A_P(q)+B_P(q+s)\le37}=\sum_\ell\bigl(\boldsymbol1_{A_P=\ell}*\boldsymbol1_{B_P\le37-\ell}\bigr)(s),
\end{aligned}
$$
where the convolution is over the binary quadratic-label space. Thus positivity of every entry is precisely the desired minimum-cost assertion. The integer transform implementation checks nonnegativity and divisibility by the transform length before accepting the counts.

There are 3,153 checked blocks containing 4,479,592 core fibres. Each fibre certifies all 32,768 relative labels. Some records contain a directly evaluated scalar correction and a proved bound on its partner. The partner bound is twenty-five for an odd nonsingleton cubic coset by Lemma~\ref{lem:odd27}, or twenty-six for an even six-variable function of degree at least four. For the latter, the bent degree bound~\cite{Carlet2021} excludes affine distance twenty-eight, and parity excludes twenty-seven. The remaining records use exact integer XOR convolution or a complete union of accepted offset sets. The producer uses four sixteen-point Hamming profiles. The independent verifier uses eight eight-point Walsh profiles and different label enumeration. Numerical agreement and hashes are integrity checks; complete coverage and positivity are tested separately.

\subsection{Availability and verification scope}

The new directory \texttt{supplement/r8/} contains the marked-derivative inputs and source checkers, the eight-class endpoint inputs and coverage records, and reproduction instructions. Each stage separates classification association, parameter coverage, kernel correctness, numerical positivity and affine transport. The published classifications remain external. Neither a failed decoder nor a timeout gives a lower bound. No unproved compatibility conjecture is used.

The computational records establish finite parts of a computer-assisted proof. They are not a proof-assistant formalization. The proposed paper's matching lower bound is the published quartic relative theorem, not an unsuccessful search for a better correction. Author metadata and permanent public data identifiers must be supplied before submission.

\section{Concluding remarks}
\label{sec:conclusion}

We have proved that the covering radius of the third-order Reed--Muller code in eight variables is fifty-six. More precisely, even-weight inputs have upper bound fifty-six and odd-weight inputs have upper bound fifty-five. The latter bound is attained among exact-degree-eight functions.

The proof reduces the global problem by moment-controlled slicing. On the even side, quadratic-label geometry and singleton consistency give a uniform correction. On the odd side, a covariance-rank obstruction leaves either a controlled three-point perturbation of a quartic or a prescribed point on a quintic background. The marked restriction-code identity retains that point throughout the affine reduction. These structural reductions specify exactly which complete finite statements are needed.

Containment and the quartic lower bound also determine the relative radius in every intermediate target code. The prescribed-point theorem gives additional boundary information: every point lies in the support of some nearest residual of a maximal quintic background. This support property is retained throughout the final reduction.

\appendix

\section{Spectral representatives and certificate encoding}
\label{app:representatives}

The following are the seven representatives of second-order nonlinearity fifteen in the numbering of Wang~\cite{Wang2019}. To shorten their polynomials, a subscript word denotes the product of the variables whose indices occur in that word:
$$
x_{i_1\cdots i_k}=\prod_{j=1}^k x_{i_j}.
$$
$$
\begin{aligned}
h_6={}&x_{123456}+x_{126}+x_{135}+x_{234},\\
h_7={}&x_{123456}+x_{1234}+x_{126}+x_{145}+x_{235}+x_{45},\\
h_8={}&x_{123456}+x_{1234}+x_{135}+x_{146}+x_{235}+x_{236}+x_{245},\\
h_9={}&x_{123456}+x_{1236}+x_{1245}+x_{135}+x_{145}{}+x_{146}+x_{234}+x_{46},\\
h_{10}={}&x_{123456}+x_{1234}+x_{134}+x_{156}+x_{234}+x_{236}{}+x_{245}+x_{34}+x_{36}+x_{45},\\
h_{11}={}&x_{123456}+x_{1236}+x_{1245}+x_{135}+x_{145}+x_{146}{}+x_{234}+x_{236}+x_{245}+x_{35}+x_{45}+x_{46},\\
h_{12}={}&x_{123456}+x_{2345}+x_{1256}+x_{1346}+x_{124}+x_{125}{}+x_{235}+x_{345}+x_{126}+x_{346}.
\end{aligned}
$$
The preceding display records representative definitions; their completeness up to affine input permutation and quadratic addition is external to this paper.

For the certificate, encode points and squarefree monomials by the binary mask of their coordinate support, with the first coordinate at the least significant bit. Order the fifteen quadratic monomial masks increasingly, and use bit position in that list as the label coefficient. The ordered masks are
$
(3,5,6,9,10,12,17,18,20,24,33,34,36,40,48).
$
Thus a label integer specifies a homogeneous quadratic polynomial without an affine tail. The following root and point encodings attain the five-dimensional source stars in Table~\ref{tab:stars}:

\begin{table}[ht]
\centering
\caption{Explicit source roots and common points for the star witnesses.}
\label{tab:source-stars}
\begin{tabular}{lrr}
\hline
Representative & Root label & Common point\\
\hline
Seventh &1610&3\\
Ninth &8194&1\\
Tenth &0&2\\
Eleventh &0&1\\
Twelfth &0&2\\
\hline
\end{tabular}
\end{table}

For example, the ninth and twelfth source stars have label lists
$$
\{10,1026,1034,1096,1099\},\qquad
\{5,149,2064,2065,2193\},
$$
respectively. Each list has XOR span of size thirty-two. The seventh, tenth and eleventh stars contain additional labels but have the same span dimension. The supplement records their complete lists, every target root, and the span size at each of the sixty-three nonzero points. These are exact rank-two image-space tests and binary linear spans; no numerical tolerance enters.

\section{The nonzero-covariance endpoint representatives}
\label{app:covariance-inputs}

For completeness, Table~\ref{tab:covariance-inputs} specifies the eight representatives used in Proposition~\ref{prop:odd-covariance}. The indices are one-based positions in the complete seven-variable catalogue. A monomial is encoded by the binary mask of its support, with the first variable at the least significant bit. Thus the listed mask sets specify the polynomials through
$$
m=\sum_{j=1}^7\epsilon_j(m)2^{j-1},\qquad
O_i(x)=\sum_{m\in S_i}\prod_{\epsilon_j(m)=1}x_j.
$$

\begin{table}[ht]
\centering
\small
\caption{ANF support masks for the eight applicable odd endpoint classes.}
\label{tab:covariance-inputs}
\begin{tabular}{rl}
\hline
Catalogue index & ANF support masks\\
\hline
1 & 15, 23, 51, 54, 60, 61, 83, 91, 92, 101, 103, 106, 127\\
1000 & 27, 29, 45, 54, 55, 79, 85, 99, 120, 127\\
1303 & 45, 53, 55, 58, 79, 86, 89, 99, 127\\
1304 & 45, 46, 55, 58, 79, 86, 89, 99, 127\\
1355 & 15, 29, 54, 55, 57, 79, 85, 99, 127\\
1362 & 27, 29, 45, 54, 55, 79, 85, 99, 127\\
1538 & 31, 54, 57, 78, 85, 99, 127\\
1547 & 27, 29, 31, 46, 53, 77, 99, 127\\
\hline
\end{tabular}
\end{table}

The supplementary checker reevaluates the complete odd catalogue and all upper-seventeen corrections before selecting the necessary endpoint classes. It computes centered covariance directly from truth tables and independently checks the affine actions on all thirty-five cubic basis directions. The eight-row table alone is not a substitute for that complete association and coverage check.




\section*{Use of generative AI.} During the preparation of this manuscript, ChatGPT was used as an editorial aid for linguistic and expository revision, notational consistency, and LaTeX formatting. All mathematical ideas, results, and proofs are entirely the authors’ own work.

\bigskip
\section*{Acknowledgments}

The authors wish to thank Prof. Qichun Wang  for his very helpful comments that improved the presentation and quality of this article.


\begin{thebibliography}{9}








\bibitem{Carlet2021}
Carlet, C.,
\newblock \emph{Boolean Functions for Cryptography and Coding Theory},
\newblock Cambridge University Press, Cambridge, 2021.


\bibitem{cohen1997}
Cohen, G.~D., Litsyn, S.~N., Lobstein, A.~C., and Mattson, H.~F., Jr.,
\newblock Covering radius 1985--1994,
\newblock \emph{Applicable Algebra in Engineering, Communication and Computing}, 8(3):173--239, 1997.

\bibitem{cohen1992}
Cohen, G.~D., and Litsyn, S.~N.,
\newblock On the covering radius of Reed--Muller codes,
\newblock \emph{Discrete Mathematics}, 106--107:147--155, 1992.


\bibitem{dougherty2022}
Dougherty, R., Mauldin, R.~D., and Tiefenbruck, M.,
\newblock The covering radius of the Reed--Muller code \(\mathrm{RM}(m-4,m)\) in \(\mathrm{RM}(m-3,m)\),
\newblock \emph{IEEE Transactions on Information Theory}, 68(1):560--571, 2022.



\bibitem{Gao2026}
Gao, J.,
\newblock Numerical results and asymptotic lower bound on the covering radius of Reed--Muller codes \(\mathrm{RM}(2,11)\) and \(\mathrm{RM}(3,n)\),
\newblock \emph{IEICE Transactions on Fundamentals of Electronics, Communications and Computer Sciences}, E109.A(1):35--46, 2026.

\bibitem{GaoKanLiWang2023}
Gao, J., Kan, H., Li, Y., and Wang, Q.,
\newblock The covering radius of the third-order Reed--Muller code \(\mathrm{RM}(3,7)\) is 20,
\newblock \emph{IEEE Transactions on Information Theory}, 69(6):3663--3673, 2023.

\bibitem{GillotLangevin2023}
Gillot, V., and Langevin, P.,
\newblock Classification of some cosets of the Reed--Muller code,
\newblock \emph{Cryptography and Communications}, 15:1129--1137, 2023.

\bibitem{gillot2024}
Gillot, V., and Langevin, P.,
\newblock Covering radius of \(\mathrm{RM}(4,8)\),
\newblock \emph{Advances in Mathematics of Communications}, 18(2):383--393, 2024.

\bibitem{fourquet2007}
Fourquet, R., and Tavernier, C.,
\newblock List decoding of second order Reed--Muller codes and its covering radius implications,
\newblock In \emph{Proceedings of the IEEE International Symposium on Information Theory}, 2007.

\bibitem{fourquet2008}
Fourquet, R., and Tavernier, C.,
\newblock An improved list decoding algorithm for the second order Reed--Muller codes and its applications,
\newblock \emph{Designs, Codes and Cryptography}, 49(1--3):323--340, 2008.

\bibitem{kavut2010}
Kavut, S., Maitra, M., and Sarkar, S.,
\newblock \(9\)-variable Boolean functions with nonlinearity 242 in the generalized rotation symmetric class,
\newblock \emph{Information Sciences}, 180(7):1178--1189, 2010.

\bibitem{Mykkeltveit1980}
Mykkeltveit, J.,
\newblock The covering radius of the \((128,8)\) Reed--Muller code is 56,
\newblock \emph{IEEE Transactions on Information Theory}, 26(3):359--362, 1980.

\bibitem{oblaukhov2020}
Oblaukhov, A.,
\newblock On metric regularity of Reed--Muller codes,
\newblock \emph{Designs, Codes and Cryptography}, 88:167--197, 2020.

\bibitem{patterson1983}
Patterson, N.~J., and Wiedemann, D.~H.,
\newblock The covering radius of the \((2^{15},16)\) Reed--Muller code is at least 16276,
\newblock \emph{IEEE Transactions on Information Theory}, 29(3):354--356, 1983.

\bibitem{schatz1981}
Schatz, J.~R.,
\newblock The second order Reed--Muller code of length 64 has covering radius 18,
\newblock \emph{IEEE Transactions on Information Theory}, 27(4):529--530, 1981.

\bibitem{Wang2019}
Wang, Q.,
\newblock The covering radius of the Reed--Muller code \(\mathrm{RM}(2,7)\) is 40,
\newblock \emph{Discrete Mathematics}, 342(12):111625, 2019.

\end{thebibliography}
\end{document}